\documentclass [reprint,prd,aps,superscriptaddress, nofootinbib,showpacs]{revtex4-1}

\usepackage[dvips]{graphicx}
\usepackage{amssymb}

\def\beq {\begin{equation}}
\def\eeq {\end{equation}}
\def\bea {\begin{eqnarray}}
\def\eea {\end{eqnarray}}

\def\nn {\nonumber}
\def\vs {\vspace}

\def\lp {\left( }
\def\rp {\right) }
\def\lb {\left[ }
\def\rb {\right] }
\def\lc {\left\{ }
\def\rc {\right\} }
\def\ra {\;\rangle }
\def\la {\langle\; }

\def\rm {\right|}
\def\rar {\rightarrow}
\def\lrar {\leftrightarrow}

\def\ub {\bar{u}}
\def\sb {\bar{s}}
\def\qb {\bar{q}}

\def\sb {\bar{s}}

\def\Kb {\bar{K}}
\def\Lb {\bar{L}}

\def\Ob {\bar{\Omega}}

\def\Tb {\bar{T}}

\def\Lv {\breve{L}}

\def\dkpp {$D^+ \! \rar \! K^- \p^+ \p^+$ }

\def\sp {\!+\!}
\def\sm {\!-\!}

\def\so {{_{1/2}}}
\def\st {{_{3/2}}}

\def\cK {{\cal{K}}}

\def\cO {{\cal{O}}}

\def\a {\alpha }
\def\b {\beta}
\def\d {\delta}
\def\D {\Delta}
\def\e {\epsilon}
\def\k {\kappa}

\def\g {\gamma}
\def\l {\lambda }
\def\L {\Lambda }

\def\p {\pi}
\def\P {\Pi}
\def\s {\sigma}

\def\th {\theta}
\def\T {\Theta}

\def\bq {\mbox{\boldmath $q$}}

\begin{document}

\title{Towards three-body unitarity in $D^+ \rar K^- \p^+ \p^+$}

\author{ P. C. Magalh\~{a}es}
\email[]{patricia@if.usp.br}
\author{M. R. Robilotta}
\affiliation{ Instituto de F\'{\i}sica, Universidade de S\~{a}o Paulo,  S\~{a}o Paulo, SP, Brazil, 05315-970;}
\author{  K. S. F. F. Guimar\~{a}es}
\author{T. Frederico}
\email[]{tobias@ita.br}
\author{ W. de Paula}
\affiliation{ Instituto Tecnol\'ogico de Aeron\'autica, 
 S\~ao Jos\'e dos Campos, SP, Brazil, 12.228-900;}
 \author{I. Bediaga}
 \author{ A. C. dos Reis}
 \email[]{alberto@cbpf.br}
\affiliation{Centro Brasileiro de Pesquisas F\'{\i}sicas, Rio de Janeiro, RJ, Brazil, 22290-180;}
 \author{ C.M. Maekawa}
 \affiliation{ Instituto de Matem\'atica, Estat\'{\i}stica e F\'{\i}sica, Universidade Federal do Rio Grande, Rio Grande, RS, Brazil; Campus Carreiros, PO Box 474, 96201-900;} 
 \author{ G. R.S. Zarnauskas}
 \affiliation{Z\"urich, Switzerland. }


\date{\today }

\begin{abstract}

We assess the importance of final state  interactions in $D^+ \rar
K^- \p^+ \p^+$, stressing the consistency between two- and
three-body interactions. The basic building block in the calculation
is a $K\pi$ amplitude based on unitarized chiral perturbation theory
and with parameters determined by a fit to elastic LASS data. Its
analytic extension to the second sheet allows the determination of
two poles, associated with the $\k$ and the $K^*(1430)$, and a
representation of the amplitude based on them is constructed. The
problem of unitarity in the three-body system is formulated in terms
of an integral equation, inspired in the Faddeev formalism, which
implements a convolution between the weak vertex and  the final
state hadronic interaction. Three different topologies are
considered for the former and, subsequently, the decay amplitude is
expressed as a perturbation series.  Each term in this series
is systematically related to the previous one and a re-summation was
performed. Remaining effects owing to single and double rescattering
processes were then added and results compared to FOCUS data.
We found that proper three-body effects are important  at threshold
and fade away rapidly at higher energies. 
 Our model, based on a vector weak vertex,
can describe qualitative features of the modulus of the decay
amplitude and agrees well with its phase in the elastic region.

\end{abstract}

\pacs{13.25.Ft 11.80.Jy 13.75.Lb}

\maketitle

\section{motivation}

About forty years ago, reactions of the type $KN \to \pi KN$ were used
to determine the $K\pi$ amplitude \cite{LASS}. Such reactions involve the
scattering of an incoming kaon and a pion from the nucleon cloud. 
The dominant
one-pion exchange amplitude is isolated by selecting events with 
low momentum
transfer. This is the only $K\pi \to K\pi$ scattering data, collected 
in the range
$0.825 < m_{K\pi} < 1.960$ GeV/$c^2$.

In the past decade, heavy flavor decays, in particular decays of 
$D$ mesons,
became a key to the physics of the light scalars. Currently these 
are the only
process in which S-wave amplitudes can be continuously studied, 
starting from
threshold, filling the existing gaps on the scattering data. 
In addition,  very
large, high purity samples, in which the initial state has always 
well defined
quantum numbers, became available in the past few years. Multibody 
decays
of heavy flavor particles proceed almost entirely via intermediate 
states involving
resonances that couple to $\pi\pi$ and $K\pi$. The universal 
$K\pi$ and $\pi\pi$
amplitudes are, therefore, present in these decays as well. 
These amplitudes could,
in principle, be extracted with great precision.

Most of the existing results come from  hadronic decays of $D$
mesons. The  golden modes  are the 
$D^+,D^+_s \to \pi^-\pi^+\pi^+$\cite{sigma}
and, in the case of the $K\pi$ system, the 
$D^+ \to K^-\pi^+\pi^+$\cite{kappa}
decay. These decay modes share some common features: the presence of
two identical particles in the final state, and a largely dominant
S-wave component. The standard procedure is the analysis of the
Dalitz plot, in which the decay amplitude is modeled by a coherent
sum of resonant amplitudes, accounting for the possible intermediate
states - the so-called isobar model. The extraction of the resonance
parameters and  decay fractions, however, depends strongly on
the particular model used for the S-wave.

The situation concerning the experimental results on the $K\pi$
amplitude is intriguing. The S-wave amplitude can be also studied
with semi-leptonic decays. In principle, in decays such as $D \to
K\pi l \nu_l$ or $\tau \to K\pi\nu_{\tau}$, the extraction of the
S-wave would be simpler than in the case of hadronic decays, for the
$K\pi$ system is free from  final state interactions (FSIs)
with the lepton pair. The $K\pi$ S-wave, in this case, should match
that of LASS\cite{LASS}, provided no energy-dependent phase is inherited 
from
the weak decay vertex. The results are conflicting, though. While
the  BaBar analysis of the decay  $D^+ \to K^-\pi^+ e^+ \nu_e$
\cite{AMO} and the FOCUS analysis of the decay $D^+ \to K^-\pi^+
\mu^+ \nu_{\mu}$ \cite{massa} conclude that the $K\pi$ amplitude is
consistent with the LASS results, the analysis of the decay  $\tau^-
\to \overline{K^0}\pi^-\nu_{\tau}$, carried out by the BaBar and
Belle Collaborations \cite{babar-tau,belle-tau} showed that these
data cannot be described by the LASS amplitude.

In 2006 the E791 Collaboration published a model independent 
analysis of the $K^-\pi^+$ S-wave amplitude using the $D^+ \to
K^-\pi^+\pi^+$ decay \cite{brian}. A very similar analysis was
performed by the FOCUS Collaboration \cite{FOCUS}. The CLEO-c
Collaboration also studied this decay, but with a somewhat different
method \cite{cleoc}. In the E791 approach, the S-wave
$K^-\pi^+$ amplitude is represented by an unknown complex function
of the $K^-\pi^+$ mass, to be determined directly from the data. The
P- and D-wave  were parameterized by the usual sum of
Breit-Wigner amplitudes. The $K^-\pi^+$ mass spectrum was uniformly
divided into 40 bins. In each bin the S-wave amplitude was defined
by two real numbers, $\mathcal{A}_0 (m^j_{K\pi}) = a_0^j
e^{i\delta^j}$. The set of 40 pairs ($a_0^j,\delta^j$) (80 free
parameters) define the S-wave through the entire $K\pi$  spectrum.
The phase and magnitude of the S-wave at an arbitrary position 
were obtained by a cubic spline interpolation.

The $K^-\pi^+$ S-wave amplitude obtained  by E791 and FOCUS is
significantly different from that from LASS.
Possible explanations for  this discrepancy fall into two broad
categories and the first one concerns the weak vertex. In the
hadronic scale, the mass of the $D$ is far both from chiral and
heavy quark limits, and the treatment of its decays cannot be
simplified by the use techniques developed in these
realms\cite{weak}. The second class of effects concerns strong
interactions, which  do take place after the weak decay. The
treatment of this part of the problem is necessarily involved since,
even in its simplest version, these FSIs already involve three
bodies  (see e.g. \cite{Azimov}). Other examples of the importance of final state interactions 
in three body decays of heavy mesons can be found in \cite{caprini, christoph, garner} As the  final mesons are 
light and have high energies, kinematics is fully relativistic and 
techniques developed in  low-energy Nuclear Physics, for the 
treatment of three-body systems,  do not apply. In the case of 
relativistic Particle Physics, in spite of a growing 
literature\cite{lc09,Zhou}, the corresponding techniques are still 
in the want of being  developed for the application to the decay problem.
Approaches to the relativistic few-body systems (two and three-body) 
have been collected
in a series of works that presents the main developments done so far
(see references 
\cite{Stadler,Polyzou,Biernat,Fred11,Carbonell,Sauli,Dorkin}).
The weak vertex  of  $D^+ \!\rar\! K^-
\p^+ \p^+$ was treated by means of  factorization and form factors,
supplemented by two-body final state interactions\cite{DiogoRafael, Ignacio}.

In this work we concentrate on the three-body structure of
strong dynamics of FSIs and postpone a detailed discussion of the
weak vertex. As one knows little about this problem, our aim is to
identify leading effects and a number of simplifications are made.
Hopefully, once these leading effects are understood, corrections
can be included in  a systematic way.

Our discussion of  three-body FSIs is based  on the theoretical
framework provided by chiral symmetry in the $SU(3)$ sector to
describe the two-meson scattering amplitude. At low energies,
chiral perturbation theory (ChPT) yields the most reliable
representation of QCD \cite{GL, EGPR}. In the present problem,
Dalitz plots for \dkpp involve energies in the range $0.4 \leq
M_{i2}^2 \leq 3\,$GeV, where the K is taken as particle 2, adopting
PDG\cite{PDG} convention. The low-energy end of this range is
directly within the scope of ChPT and it is in this region where
discrepancies between LASS\cite{LASS} and FOCUS\cite{FOCUS} data are
more pronounced. Even outside this range, the chiral framework still
proves to be very useful and provides rigorous guidance for possible
extensions.

The case of the $\k$ is paradigmatic. This state was found in a
number of experiments\cite{brian,FOCUS, cleoc, kappa, BES} and
theoretical models\cite{kappaTeoria, OO, re-Kpi,Mussalam}, with a
complex mass $m_\k = [(0.672 \mp 0.040) \sm i\,(0.550 \sm
0.034)]\;$GeV\cite{PDG}, which can be considered as low in the
hadronic scale. For this very reason, it definitely cannot be
accommodated as a fundamental resonance in a chiral
lagrangian\cite{EGPR}. In ChPT, resonances can only be introduced as
next-to-leading order, which decay by means of
explicit couplings to pseudoscalar mesons. In this framework,
a resonance should become a stable particle when its coupling to the
$K\pi$ system is turned off.

The $\k$ does not fall in this category. Its relationship with ChPT
is a more subtle one and has been clarified about a decade
ago\cite{OO}, with the realization that unitarization of low-energy
chiral results gives rise to amplitudes which have poles in the
complex $s$-plane. The $\k$ corresponds to the state with the lowest
energy and it is present even if one considers only leading order
contact interactions (see, for instance, our fig.\ref{FFCNP}). This
idea proved to be very fruitful and motivated both reanalysis of
low-energy experimental $K\pi$ data\cite{re-Kpi,Mussalam} and the
interpretation of the $\k$ as a tetraquark or two-meson
state\cite{Pelaez}. The case of the $\s$-meson, which occurs in
$SU(2)$, is totally similar ( see e.g. \cite{Tetraquarks, maiani, wayne}). There, however, the availability of
precision data allows the pole to be extracted directly from the 
$\p\p$ amplitude, in a model independent way\cite{sigma-pole}.

Our presentation is divided as follows. In sect.II we review, for
the sake of completeness, chiral results for the $K\p$ amplitude and
its unitarization. The conceptual framework for the three-body
unitarization is introduced in sect.III and,  in sect.IV, its
perturbative expansion is used to derive predictions for the $D^+
\rar K^- \p^+ \p^+$ amplitude. Results and conclusions are given in
sect.V. The manuscript also includes several appendices
dealing with technical matters.

\section{$\Kb\p$ amplitude}

\subsection{Interaction kernel}

 The reaction  $\p_a(p_a)\,\Kb_b(p_b) \rar \p_c(p_c) \, \Kb_d(p_d)$
is described in terms of the usual Mandelstam variables
$s, t, u$, constrained by the condition
$s + t + u = p_a^2 + p_b^2 + p_c^{2} + p_d^{2}\;$.
In the  CM, results can be expressed in terms of the momentum $\bq$,
with
$\bq^2 = s\, \rho^2/4$,
$ \rho = \sqrt{1 - 2\,(M_\p^2 \sp M_K^2)/s + (M_\p^2 \sm M_K^2)^2/s^2}$.

\begin{figure}[h]
\includegraphics[width=0.7\columnwidth,angle=0]{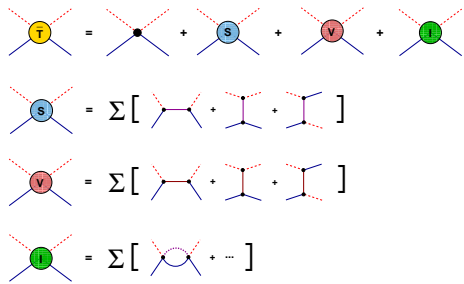}
\caption{ $K\p$ amplitude: the contact diagram is the leading one at 
low-energies, 
whereas blobs are corrections due to other intermediate states; $S$ 
and $V$ correspond
to scalar and vector resonances and $I$, to inelastic channels; the 
summation signs 
indicate the possibility of more than one intermediate state of each kind. }
\label{fig:1}
\end{figure}

Chiral perturbation theory determines the tree amplitudes $\Tb_I$,
with isospin $I$, as the sum of a $\cO(q^2)$ contact
term\cite{GL}, supplemented by $\cO(q^4)$ scalar and vector
resonances\cite{EGPR}, together with inelastic contributions\cite{re-Kpi}.
The amplitude for each channel, indicated in fig.\ref{fig:1}, is then 
written as
\bea
&& \Tb_I = \Tb^c + \Tb^S  + \Tb^V + \Tb^I  \;.
\label{ea.1}
\eea
%

In this exploratory work, we are interested in determining
dominant structures in the scalar sector and, in the sequence, we neglect
vector resonances and inelasticities.
 As shown in fig.1, scalar resonances contribute to $s$, 
$t$ and $u$ channels. 
Each of the corresponding amplitudes has the form 
$\Tb^S(x)=\a_x(x)/(x \sm m_x^2)\,$, 
where $x=s,t,u\,$, $\a_x$ is a $\cO(q^4)$ polynomial and $m_x\,$, a large
resonance mass. 
In the physical region, the variables $t$ and $u$ are negative,
whereas $s$ is positive.
This means that $\Tb^S(s)$ can become arbitrarily large, whereas
$\Tb^S(t)$ and $\Tb^S(u)$ remain always finite.
Close to threshold, on the other hand, all the $\Tb^S(x)$ are smaller,
by construction, than the contact term $\Tb^c\,$.
Therefore, in the entire physical region, the combination
$[\Tb^S(t)+\Tb^S(u)]$ can be neglected in comparison with
$[\Tb^c + \Tb^S(s)]\,$. 
We thus identify the leading tree $K\p$ amplitudes with the functions
\bea
\Tb_\so &\!=\!&
\frac{1}{F^2} \,\lb s + 3\, t/4 - (M_\p^2 \sp M_K^2)\rb
- \sum_i^N \;\frac{\a_i(s)}{s \sm m_i^2} \;, 
\label{ea.2a}\\[2mm]
\a_i &\!=\!&  \frac{3}{2F^4} \;
[c_d^i\,s - (c_d^i \sm c_m^i) \, (M_\p^2 \sp M_K^2)\,]^2 \;,
\label{ea.2}\\[4mm]
\Tb_\st^c &\!=\!&
- \,\frac{1}{2 F^2} \lb s - (M_\p^2 \sp M_K^2)\rb \;,
\label{ea.3}
\eea

where $F$ is the meson decay constant, $N$ is the number of scalar
resonances of the  $J^P =0^+$ considered, with masses $m_i$
and coupled to two mesons by the constants $c_d^i$ and $c_m^i$, 
estimated in ref.\cite{EGPR}.
Their values are such that $c_d \sim c_m$ and
$(c_d\,c_m/m_i^2) \sim 10^{-3}\,$,
explaining why $\Tb^S(t)$ and $\Tb^S(u)$ are much smaller than 
the terms kept in eq.(\ref{ea.2a}). The neglect of these terms allows the $K\p$
amplitude to be unitarized in a very compact form, as discussed in the sequence.
The $I=3/2$ amplitude is repulsive, contributes little to FSIs
and is also neglected in the sequence.
Projecting out the $S$-wave, one finds the leading kernel of the
dynamical equation, given by
\bea
\cK_{S \so} \equiv \cK &\!=\!& \cK_c
- \sum_i^N  \frac{\a_i(s)}{s \sm m_i^2} \;,
\label{ea.4}\\[2mm]
\cK_c &\!=\!&
\lb 5 s/8 - (M_\p^2 \sp M_K^2)/4 \right. 
\nn \\&&-\left. 3\,(M_\p^2 \sm M_K^2)^2/8s \rb /F^2
\;.
\label{ea.5}
\eea

This kernel is real and hence suited for describing elastic
processes only. The inclusion of inelasticities, due mostly to 
intermediate states containing $\eta_1$ and $\eta_8$, can be performed
by means of well known coupled-channels techniques\cite{re-Kpi}.
We remain within the elastic approximation
and derive $T$, the elastic $I=1/2$, $S$-wave $\p K$ scattering amplitude,
by means of the Bethe-Salpeter equation (BSE),
written schematically as
\bea
T(s)&=& \cK(s)
- i \int \frac{d^4 \ell}{(2\p)^4}\; \cK(p,\ell) \;
\frac{1}
{[(\ell \sp p/2)^2 \sm M_\p^2 \sp i\,\e\,]}\;  \nn \\
&&\times \, \frac{T(p,\ell) }{[(\ell \sm p/2)^2 \sm M_K^2 \sp i\,\e\,]}
 \; \;,
\label{ea.6}
\eea
with $p^2 \!=\! s$.
Our approximations ensure that both $\cK$ and $T$
do not depend on $\ell$ and the BSE acquires the very simple form
\beq
T = \lb 1 - T\; \Omega \rb \; \cK \;,
\label{ea.7}
\eeq
where the two-meson propagator is
\bea
\Omega(s) &=& i\, \int \frac{d^4 \ell}{(2\p)^4}\;
\frac{1}
{[(\ell \sp p/2)^2 \sm M_\p^2 \sp i\,\e\,]}\; \nn\\
&& \times \,\frac{1}{[(\ell \sm p/2)^2 \sm M_K^2 \sp i\,\e\,]} \;.
\label{ea.8}
\eea
This integral is ultraviolet divergent, but this unwanted behaviour
can be cured by means of a subtraction.
Following ref.\cite{GL}, we write the regular part of $\Omega$ as
\bea
&& \Ob(s)= \Omega(s)-\Omega(0)\;
\label{ea.9}
\eea
where the divergent part of $\Omega(s)$ is contained in $ \Omega(0)$.
Regularization amounts to the replacement
$\Omega(0) \rar C$, where $C$ is an unknown finite constant which
has to be fixed by experimental input.
The function $\Ob$ can be evaluated analytically, and explicit
results are given in  appendix \ref{bolha} .
After regularization, eq.(\ref{ea.7}) becomes
\beq
T = \lb 1 - T\; (C + \Ob) \rb \; \cK \;,
\label{ea.10}
\eeq
and its  solution reads
\bea 
&& T = \cK/D \;,
\label{ea.11}\\ [2mm] 
&& D = 1 + (C + \Ob) \,\cK \;.
\label{ea.11a}
\eea

The $K\p$ amplitude is thus determined by a
rather simple algebraic equation, involving just the
two-meson propagator and the kernel.
The latter, in turn, includes only contact interactions and
$s$-channel resonances.
In the sequence we will use the result
\beq
T\;(C + \Ob)= 1 - 1/D \;,
\label{ea.12}
\eeq
derived from eq.(\ref{ea.11}).
The function $\Ob$ is real below the threshold at
$s = (M_\p \sp M_K)^2$ and acquires an imaginary
component above it.
 In the physical region, below the first inelastic 
threshold, the kernel is real, whereas 
$\Ob = \Ob_R + i\, \Ob_I$, $\Ob_I = - \rho/16 \p$, and the unitary
amplitude can be represented by means of a real phase shift $\delta$ \cite{borges} as
\bea
&& T = \frac{16 \p}{\rho} \,  \sin \d \, e^{i \d}
\;\lrar \;
\tan\d = -\, \frac{\Ob_I \, \cK}{1 + (C \sp \Ob_R) \, \cK} \;.
\label{ea.13}
\eea
 For energies above inelastic thresholds,
the kernel $\cK$ acquires imaginary components owing to processes
included in the blob $I$ of fig.1 and the amplitude is damped.

\subsection{ Alternative representations}

As a side comment, we note that the representation of $T$, in terms of 
Breit-Wigner structures,
 is not suited to this problem.
They apply just to a single isolated resonance,
since eq.(\ref{ea.4}) would read $\cK=-\a_1/(s-m_1^2)$
and eq.(\ref{ea.11}) could be written as
\bea
T = \frac{-\a_1}
{s - [m_1^2 + (C+\Ob_R)\,\a_1]+i\,[\rho\,\a_1/16\p]} \;.
\label{ea.14}
\eea
In this case, the terms $[m_1^2 + (C+\Ob_R)\,\a_1]$ and $[\rho\,\a_1/16\p]$
could be identified as a running mass and a width respectively.
On the other hand, in the more realistic situation where both
the contact interaction and other resonances are present, the
coupled structure of the problem shows up.
For instance, in the case of a contact term supplemented by two
resonances, the amplitude (\ref{ea.11}) would read
\bea
T &=& \frac{(s \sm m_1^2)\,(s \sm m_2^2)\,\cK_c
\sm (s \sm m_2^2)\,\a_1 \sm (s \sm m_1^2)\,\a_2}
{ 1 \sp (C \sp \Ob_R \sp i\, \Ob_I)
[\,\cK_c
\sm \,\a_1 / (s \sm m_1^2) \sm \a_2/(s \sm m_1^2)\,]}  \nn\\
&& \times \, \frac{1}{(s \sm m_1^2)\,(s \sm m_2^2)}\;.
\label{ea.15}
\eea
The coupling of various types of interaction
gives rise to a complicated structure which cannot
be written naturally as either products or sums of individual Breit-Wigner
expressions for each resonance.

\subsection{ Data and poles - schematic features}

\begin{figure}[h]
\includegraphics[width=0.8\columnwidth,angle=0]{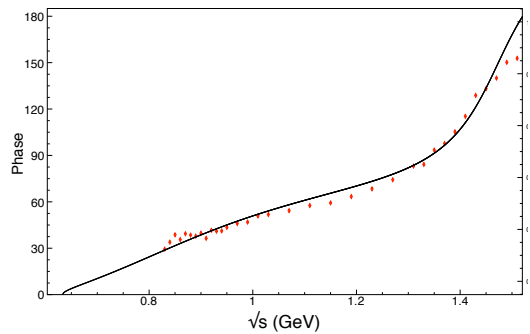}
\caption{$S$-wave isospin $1/2$ phase shift:
 LASS data\cite{LASS} and our fit.}
\label{FFIT}
\end{figure}

The main qualitative features of the FSIs can be understood
by means of a $K\p$ amplitude as given just by a chiral
contact term, supplemented by a single resonance.
When more resonances are included, these features are preserved,
but the amount of algebraic work increases considerably.
We work with the simplest version, which corresponds to
\bea
&& T = \frac{\cK}{1 + (C + \Ob) \,\cK}\;,
\label{ea.16}
\\[2mm]
&& \cK = \cK_c - \a_1/(s \sm m_1^2) \;. \nn
\eea
This amplitude depends on two sets of 3 parameters,
namely $[F, M_\p, M_K]$ and $[m_1, c_d^1, c_m^1]$,
besides the subtraction constant $C$.
The first set is determined by chiral perturbation theory
and we adopt $F=\sqrt{F_\p\,F_K}=0.102722\;$GeV\cite{re-Kpi},
$M_\p^+ = 0.1396\;$GeV, $M_K^+ = 0.4937\;$GeV.
The other one is obtained by fitting  LASS data\cite{LASS} in the
elastic region and we find $C=-1 \times 10^{-3}\,$,
$m_1= 1.330 \;$GeV, $c_d^1 = 0.0352 \;$GeV and $c_m^1= 0.001027 \;$GeV,
which yield the curve shown in fig.\ref{FFIT}.
Results for the coupling constants are close to one of the sets
given in ref.\cite{re-Kpi} $[c_d,c_m=0.030,0.043]GeV$ and
roughly consistent with those derived from the decay
$a_0 \! \rar \! \eta \p$ in ref.\cite{EGPR},
namely $c_d = 0.032\;$GeV and $c_m = 0.042\;$GeV.
%

\begin{figure}[h]
\includegraphics[width=0.6\columnwidth,angle=0]{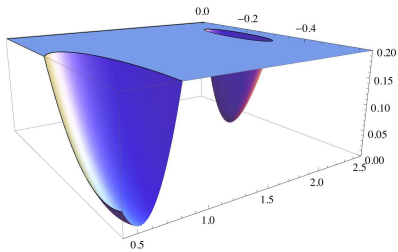}
\caption{ Two poles in $|D(s)|$ in the complex $s$ plane.  }
\label{FFCN}
\end{figure}

\begin{figure}[h]
\includegraphics[width=0.6\columnwidth,angle=0]{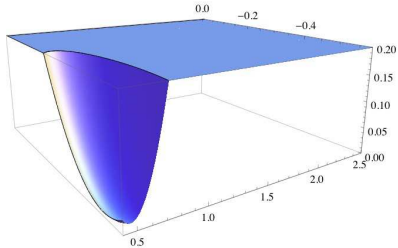}
\caption{A single pole in $|D(s)|$ in the complex $s$ plane.}
\label{FFCNP}
\end{figure}

Feeding the parameters from the fit into eq.(\ref{ea.16}) and extending
it to the second Riemann sheet, one finds two coupled poles in the complex
$s$-plane, given by the condition $D(s) = 0 $\,.
They are located at $s=\th_i$, with
$\th_\k = 0.51426725 \sm i\, 0.51423116\;$GeV$^2$
and $\th_1 = 2.16256534 \sm i\,0.24130498 \;$GeV$^2$, and identified 
respectively
with the $\k$ and the $K_0^*(1430)$.
The latter is compatible with the PDG\cite{PDG} value,
$s_{(1430)} = [(2.01 \pm 0.15) - i\, (0.38 \pm 0.13)]\,$GeV$^2$.
In order to produce a feeling for the strength of the
coupling between these resonances, we keep just the contact term in
eq.(\ref{ea.16}), by setting  $\a_1 = 0\,$,
finding $\th_\k = 0.45505779 \sm i\, 0.51167711 \;$GeV$^2$ and 
$\th_1 = m^2_1$.
The $\k$-pole is thus rather stable.
The behaviour of the function $D(s)$ along the complex $s$-plane
is shown in fig.\ref{FFCN} and it is possible to
see two wells around the $\k$ and $K_0^*(1430)$ poles.
The plot in fig.\ref{FFCNP}  corresponds to the case $\a_1 = 0\,$,
in which just the $\k$ exists, dynamically generated.

%

%

\begin{figure}[h]
\hspace*{-30mm}
\includegraphics[width=0.6\columnwidth,angle=0]{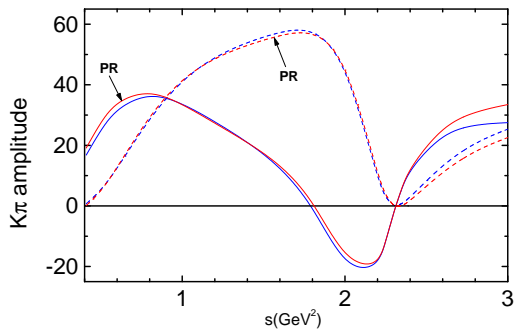}
\caption{Real (full lines) and imaginary (dashed lines) components of
the $K\p$ amplitude from eq.(\ref{ea.16})(blue) and the pole
representation (PR), eq.(\ref{ea.17}) (red).}
\label{FFamplitude}
\end{figure}

The fact that the function $D(s)$ has two poles in the
second Riemann sheet, allows the elastic amplitude,
eq.(\ref{ea.11}), to be represented as
\bea
&& T = \frac{\cK}{D}
= \cK \lb \b + \frac{\g_\k}{s-\th_\k} + \frac{\g_1}{s-\th_1} \;\rb\;,
\label{ea.17}
\eea
with
$\b = 0.2200\;$, $\g_\k = -0.1849 + i\, 0.6378 \;$GeV$^2$,
$\g_1 = 0.2247 + i\, 0.1260 \;$GeV$^2$.
In fig.\ref{FFamplitude} we compare it with the exact amplitude and learn 
that this simple representation is reasonable in the range covered by 
the Dalitz plot.
 We adopt it in the present exploratory work and leave a more complete
  treatment for a future investigation.

%

\section{Three-body unitarity}

The amplitude $D^+ \rar K^- \p^+ \p^+$, denoted by
$A(m_{12}^2,m_{23}^2)$, is symmetric under the exchange of the final
pions and written as
\bea A(m_{12}^2,m_{23}^2) &\!=\!& W(m_{12}^2,m_{23}^2) + a(m_{12}^2)
+ a(m_{23}^2)\;,
\label{tbu.1}
\eea
where $W$ is the weak vertex and the functions $a$ incorporate
hadronic FSIs.

\begin{figure}[ht]
\includegraphics[width=0.9\columnwidth,angle=0]{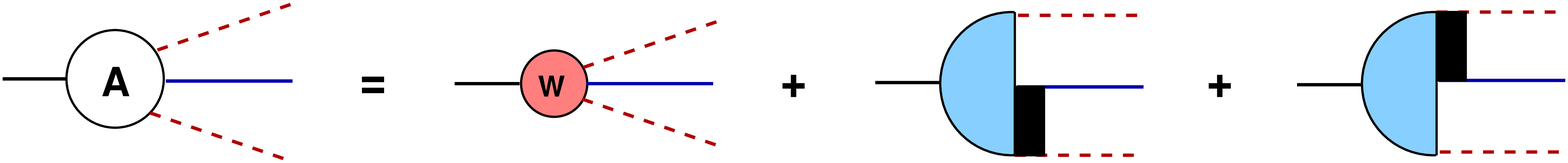}
\caption{ Diagrammatic representation of the heavy meson decay
process into $K\pi\pi$, starting from the partonic amplitude (red)
and adding hadronic multiple scattering in the ladder
approximation.}
\label{FFSI-1}
\vs{4mm}
\includegraphics[width=.99\columnwidth,angle=0]{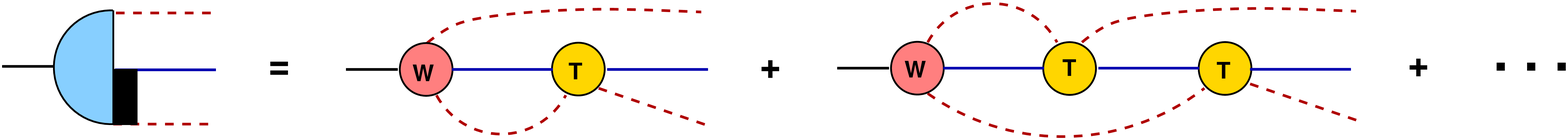}
\caption{Rescattering series implementing three-body unitary.}
\label{FFSI-2}
\end{figure}

In figs. \ref{FFSI-1} and \ref{FFSI-2} we show the structure of
$A$ in terms of the rescattering series, which depends on $T$, 
the $K\p$ amplitude obtained in the preceding
section.
The diagram involving just $W$ in  fig.~\ref{FFSI-1} 
describes the possibility
that the mesons produced in the decay reach the detector without
interacting.
 As the Fermi coupling constant $G_F$ 
entering $W$  is very small,
the amplitude $A(m_{12}^2,m_{23}^2)$ is linear in this parameter, 
indicating that each dynamical component of the weak vertex 
in fig.\ref{FFSI-1} evolves independently by means of FSIs.
In other words, if the primary vertex can be written as a sum 
$W=\sum_i\,W_{(i)}\,$ of dynamical contributions,
one automatically has $A = \sum_i\,A_{(i)}\,$.

In principle, $\p^+ \p^+$ interactions could contribute to FSIs,
but this subsystem has isospin 2 and this channel can be
safely neglected, because phase shifts are  small. 
Three-body unitarity is then dominated by the series
represented in  fig.~\ref{FFSI-2}, where the first diagram
corresponds to the
spectator-pion approximation, often found in the literature.
Contributions from three-body irreducible rescattering process are
terms of the series which contain successive
interactions among different pairs.
The second diagram is thus the lowest order contribution
of this kind.

\begin{figure}[ht]
\includegraphics[width=.99\columnwidth,angle=0]{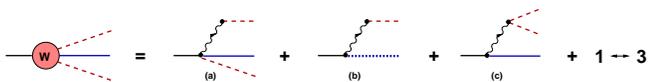}
\caption{ Topologies for the weak vertex: the dotted line is a scalar 
resonance and the wavy line is a $W^+$, which is contracted to a point 
in calculation; in diagram $c$, one of the pions is neutral.}
\label{FFSI-3}
\end{figure}

 In this work we concentrate on the perturbative structure 
of the series describing strong final state interactions.
In order to perform such a study, one needs 
information concerning the production of the $K\p\p$ system
in the primary weak decay.
A suitable conceptual point of departure is the quark diagram approach 
described by Chao\cite{chau},
based on 6 independent topologies.
In spite of their symbolic character and the absence of interactions
mediated by gluons, they implement properly the CKM quark mixing.
In the specific case of non-leptonic $D$ decays, leading contributions
are incorporated into the hadronic effective lagrangian 
given by Bauer, Stech and Wirbel\cite{BSW} 
as
\bea
L_{\mathrm{eff}} = \frac{G_F}{\sqrt{2}}
\lc a_1\,(\ub\,d')_H\;(\sb' \, c)_H 
+ a_2\,(\sb' \, d')_H\;(\ub \, c)_H \rc\;,
\label{bsw}
\eea
where $(\qb \, q)_H $ represents hadronic expectation values 
of $(V \sm A)$ charged weak currents and, the QCD factors $a1$ and $a2$ are related to the Wilson coefficients  with ratio roughly given by  $a_2/a_1 \sim -0.4\,$.
A detailed study of the process $D^+ \rar K^- \p^+ \p^+$ based
on this lagrangian has been performed by Boito and 
Escribano\cite{DiogoRafael}, considering two-body FSIs only.
These authors assessed the relative importance of contributions 
proportional to either $a_1$ or $a_2$ to observables and the latter
was found to rather visible, although systematically smaller than 
the former.
In this work we focus on the strong evolution of leading contributions  
and neglect, for the moment, the color suppressed term proportional 
to $a_2\,$.
The light quark sector in the term proportional to $a_1$
involves just $(\ub\,d)_H\,$, which is minimally realized by the 
matrix elements
$\la \p^+ | A^+ |0\ra $ and $\la \p^0\p^+ |V^+|0 \ra\,$.
Concerning the factor $(\sb \, c)_H\,$, we allow the final strange quark 
to be carried by either a kaon or a strange scalar resonance,
and arrive at the set of topologies indicated schematically in 
fig.~\ref{FFSI-3}.  
The strengths of these vertices are respectively $W_a$,
$W_b$ and $W_c$,
assumed provisionally as constants.
In the case of $W_a$, one has the isospin substructure:
$K^-(k) \; \p^+(q)\; \p^+(q') \rar \sqrt{2/3}\;\; W_a\,$ and
$\bar{K}^0(k) \; \p^0(q)\; \p^+(q') \rar - \sqrt{1/3}\;\; W_a\,$.

The construction of the amplitudes $a(m_{i2}^2)$ is discussed in the
sequence and formulated
in terms of the rather general $K\p$ amplitude given by eq.(\ref{ea.11}).
Hence it does not depend on
values adopted for parameters, the number of explicit resonances
considered and possible couplings to inelastic channels.
The function $a(m_{12}^2)$ is written as
\bea
a(m_{12}^2) &\! = \!& \sum_{N=1}^\infty a_N(m_{12}^2) \;,
\label{tbu.2}
\eea
where $N$ is the number of $K \p$ amplitudes intervening
in a particular diagram of  fig.~\ref{FFSI-2}.
The sum in (\ref{tbu.2}) can be performed by means of a Faddeev-like
decomposition of the amplitude $A(m_{12}^2,m_{23}^2)$. The Faddeev
components are identified with  $a(m_{12}^2)$ and
$a(m_{23}^2)$, which are the nonperturbative solution of a
scattering equation.

Following the model proposed in ref. \cite{lc09}, we write the full
decay amplitude of  fig.~\ref{FFSI-2} as:
%
\begin{small}
\bea
&& A(m_{12}^2,m_{23}^2) = W \lb 1-\int \frac{d^4q \, d^4q'}{(2\p )^8}
{T_{3\to3} (k, k' ;q ,q')\over
(q^2 - M^2_\p + i\, \epsilon)\; } \, 
\nonumber \right.\\&& \times \, \left.   
 \frac{1}{(q'^2-M^2_\p + i\, \epsilon)\;
[(P - q' -q)^2-M^2_K+ i\, \epsilon]} \rb,
\label{deq1}
\eea

\end{small}
 where $P$, $k$ and $k'$ are respectively the momenta of the $D$
and of the pions produced in its decay.
The matrix element of the $3\to 3$
transition matrix is $T_{3\to3}(k,k';q,q')$. In order to simplify the 
description, we use the point-like weak vertex.

The weak vertex for the decay of the $D$ meson into the $K\pi\pi$
channel is convoluted with the $3\to 3$ off-shell transition matrix,
which takes into account the three-meson interacting final state, as
shown in fig.~\ref{FFSI-2}, including the three-body connected
ladder series, where the $2\to 2$ scattering process is summed up in
the $K\pi$ transition matrix.

The $3\to 3$ transition matrix is obtained from the following
assumptions: {\it i)} the $K\pi\pi$ Bethe-Salpeter equation is
solved in the ladder approximation, and {\it ii)} the $K\pi$
transition matrix is effective in the S-wave states. The full $3 \to
3$ ladder scattering series is summed up  when the integral
equations for the Faddeev decomposition of the scattering matrix  are solved.

The matrix elements of our $K\pi$ amplitude depend just on the
Mandelstam variable $s$ of the two-body subsystem, as given by eq.(\ref{ea.7})
allowing the Faddeev components of the decay amplitude
to be written in a factorized form:
\bea
a(m_{12}^2)=T(m^2_{12}) \;\xi(p_{3}) \;.
\label{deq2}
\eea
The function $a(m_{12}^2)$ thus carries the full
effect of the final state interaction through the two-meson
amplitude $T$ multiplied by a spectator amplitude $\xi$,
which contains the full three-body rescattering contributions,
given by the sum of all processes in the ladder approximation for
the multiple scattering series.

\begin{figure}[h]
\includegraphics[width=0.9\columnwidth,angle=0]{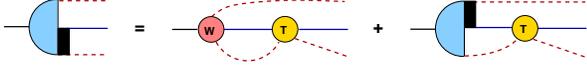}
\caption{Diagrammatic representation of the integral equation for
the three-body function $T(m_{12}^2)\xi(k_3)$ (left).
The driving term contains the partonic amplitude from the weak vertex convoluted
with the two-body scattering amplitude (right, first graph).}
\label{FFSI-4}
\end{figure}

The re-summation of the scattering series in the reduced amplitude
$\xi(k)$ can be done by an integral equation corresponding to
fig.\ref{FFSI-4}, given by
\bea
\xi(k)&=&  \xi_1(k)  - \,i \int \frac{d^4q}{(2\p)^4}
{T[(P-q)^2] ~\over
(q^2-M_\p^2+ i \epsilon)} \nn\\ && \times \,\frac{\xi(q)}{[(P \sm k \sm q)^2-M_K^2+ i \epsilon]} , \nn
\label{deq5x}
\eea
where the driving term is
\bea
\xi_1(k)= - i\, W \int \frac{d^4q}{(2\p)^4}
{1\over(q^2-M_\p^2+ i \epsilon)} \,\nn \\ \times \frac{1}{\;[(P \sm k \sm q)^2-M_K^2+ i \epsilon]}  ,
\label{deq6}
\eea
\\[-4mm]
and depends just on a one-loop integral.
This term, multiplied by $T(m_{12}^2)$,
gives rise to the first diagram in fig.\ref{FFSI-2}, whereas
the second term in eq.(\ref{deq5x}), multiplied by $T(m_{12}^2)$
represents the sum of all the remaining three-body
connected processes shown in that figure.
The lowest order connected contribution is the
second diagram in fig.\ref{FFSI-4}.

The contribution to the three-body rescattering process given by
eq.(\ref{deq5x}) is built by mixing interactions from
the two possible $K\pi$ pairs.
The resulting reduced amplitude is a function of just the
momentum of the final spectator pion.
The model separates the decay amplitude into
two parts:
one corresponding to a smooth function of the momenta of the pions in
$W$ and another given by $\xi(k)$ times the pair amplitude,
which  is a fully three-body interacting term modulated
by the $K\pi$ amplitude.

So far, we did not consider isospin degrees of freedom.
The kernel of the integral equation (\ref{deq5x}) involves the
change between the pions corresponding to the final isospin channel
of the pair $K\pi$ and gives rise to isospin factors discussed
in appendix \ref{isospin}.
The re-coupling coefficient given by eq.(\ref{recou2})
appears weighting the kernel.
Taking into account also the isospin weight for the driving term,
we find
\bea
\xi(k)&=& \frac53 \,\xi_1(k) 
-   \frac23\, i \int \frac{d^4q}{(2\pi)^4}
T[(P-q)^2]~
\over(q^2-M_\pi^2+ i \epsilon) \,\nn \\ &&\times\, \frac{\xi(q)}{[(P-k-q)^2-M_K^2+ i \epsilon]} \;. 
\label{deq5i}
\end{eqnarray}

As the main purpose of this work is to investigate the effect of the
three-body unitarity on the decay amplitude, we analyze in the
following section the perturbative contributions to the FSI at
one- and two-loop approximations. We choose to exemplify the 
series expansion of eq.(\ref{deq5i}) for the  weak vertices $a$ and $c$ of 
 fig.~\ref{FFSI-3} and present the case of vertex $b$ when discussing the perturbative 
calculation. In the case of vertex $a$, we find
\bea
A_a(m_{12}^2,m_{23}^2) &\!=\!& \sqrt{\frac{2}{3}}  W_a
\lc 1 + T(m_{12}^2) \, \frac{5}{3}\,
\lb \xi_1(m_{12}^2) 
\right.\right. \nn\\ &\!+\!&  \left.\left.
 \frac{2}{3} \; \xi_2(m_{12}^2)+
\lp\frac{2}{3}\rp^2 \xi_3(m_{12}^2) 
\cdots \rb \rc \nn\\
&+& (1 \lrar 3)\;. 
\label{pu.3x}
\eea
where the argument of the function $\xi$ is written in terms of the
invariant mass squared of the $K\pi$ subsystem, i.e.,
$m_{12}^2=(P-p_3)^2$,  instead of the individual momenta.  
The factor $ \sqrt{\frac{2}{3}}$ comes from
the isospin projection of  the $K\pi$ pair in the weak vertices to $I=1/2$. The
perturbative $n$-loop amplitude is constructed recursively as:
 \bea
\xi_{n}[(P-k)^2]= -  i \int \frac{d^4q}{(2\pi)^4}
{T[(P-q)^2]~ \over(q^2-M_\pi^2+ i
\epsilon)} \nn \\ \times \,
\frac{\xi_{n-1}[(P-q)^2]}{[(P-k-q)^2-M_K^2+ i \epsilon]} \;.
 \label{deq6i}
\end{eqnarray}

For later convenience we introduce the function
$\lambda_{n}(m^2_{12})$, defined as
\begin{equation}
\lambda_{n}(m^2_{12})=T(m^2_{12})\xi_{n} (m^2_{12}) \ , \label{deq6ii}
\end{equation}
which is useful within our approximation of disregarding the
momentum structure of the weak vertex,
and  eq.(\ref{pu.3x}) becomes
\bea A_a(m_{12}^2,m_{23}^2) &\!=\!&\sqrt{\frac{2}{3}}\; W_a\lc 1 +
\frac{5}{3} \lb \l_1(m_{12}^2) + \frac{2}{3} \; \l_2(m_{12}^2) +
\right.\right.
\nn\\
&\!+\!& \left.\left. \lp
\frac{2}{3} \rp^2 \l_3(m_{12}^2) 
\cdots \rb \rc + (1 \lrar 3) .
 \label{pu.3}
\eea

The three-body re-scattering series starting from the weak vertex
$b$ has to be treated properly in order to avoid double counting in
the scattering series in the two-meson channel, as the scalar
resonance is dressed by the $K\pi$ interaction  (c.f. fig.
\ref{FPU-3}). In the case of vertex $c$ the scattering series
simplifies as the $\pi_0$ produced directly from the $W$ decay is
not present in the final state  and it is written as:
 \bea A_c(m_{12}^2,m_{23}^2)
&\!=\!&-\frac{\sqrt{2}}{3}\; W_c\lc   \l_1(m_{12}^2)
+ \frac{2}{3} \; \l_2(m_{12}^2)
\right.
\nn\\[2mm]
&\!+\!& \left.
  \lp \frac{2}{3} \rp^2
\l_3(m_{12}^2)  \cdots \rc + (1 \lrar 3)\;.
 \label{pu-c.3}
\eea
%

\section{ Perturbative processes}

\begin{figure}[h]
\includegraphics[width=0.5\columnwidth,angle=0]{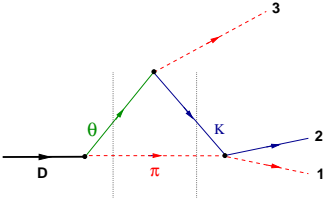}
\caption{Triangle diagram representing the reaction
$D(P) \rar [\p K](p_{12})\; \p(p_3)$, with intermediate states
$\p(p_{12} \sm k)$, $\th(p_3 \sp k)$, $K(k)$,
associated with the integral given in eq.(\ref{pu.1});
invariant masses along the dotted lines can be smaller than those
of external lines.}
\label{FPU-1}
\end{figure}

In this section, the first two terms of the function
$a(m_{12}^2)$ given by eq.(\ref{tbu.2}) are evaluated covariantly and different contributions are classified according to the
type of initial weak vertex.
Diagrams involve two kinds of loops, containing either two or three meson propagators.
The former require regularization and are treated as in the
construction of the $K\p$ amplitude presented in  sect. II.
The latter are {\em triangle integrals}, written as
\bea
I_{\p K \th}(m_{12}^2) &=&
\int \! \frac{d^4 k}{(2\p)^4} \;
\frac{1}{[(p_{12} \sm k)^2 \sm M_\p^2 \sp i\,\e]\;} 
\nn\\
&& \times \,\frac{1}{[[k^2 \sm M_K^2 \sp i\,\e]\;
(p_3 \sp k)^2 - \th\,]} \;,
\label{pu.1}
\eea
where $\th=\th_R - i\,\th_I$, is the position of the pole in the
complex $s$-plane, with $\th_R$ and $\th_I$ constant positive quantities.
This integral is similar to those occurring  in usual
calculations of form factors, but not identical,
since the invariant masses along the dotted lines in
fig.\ref{FPU-1} can be smaller than either $m_D^2$ or $m_{12}^2$.
It is thus mathematically more akin to integrals needed to
describe form factors of unstable particles, such as the $\D$,
$\rho$ or $K^*$.
We write
\bea
I_{\p K \th} &\!=\!& i\,\P_{\p K \th}/(4\p)^2
\label{pu.2}
\eea
and the evaluation of the functions $\P$ is discussed
in appendix \ref{TrI}.

\subsection{Contributions proportional to $W_a$}

\begin{figure}[h]
\includegraphics[width=0.9\columnwidth,angle=0]{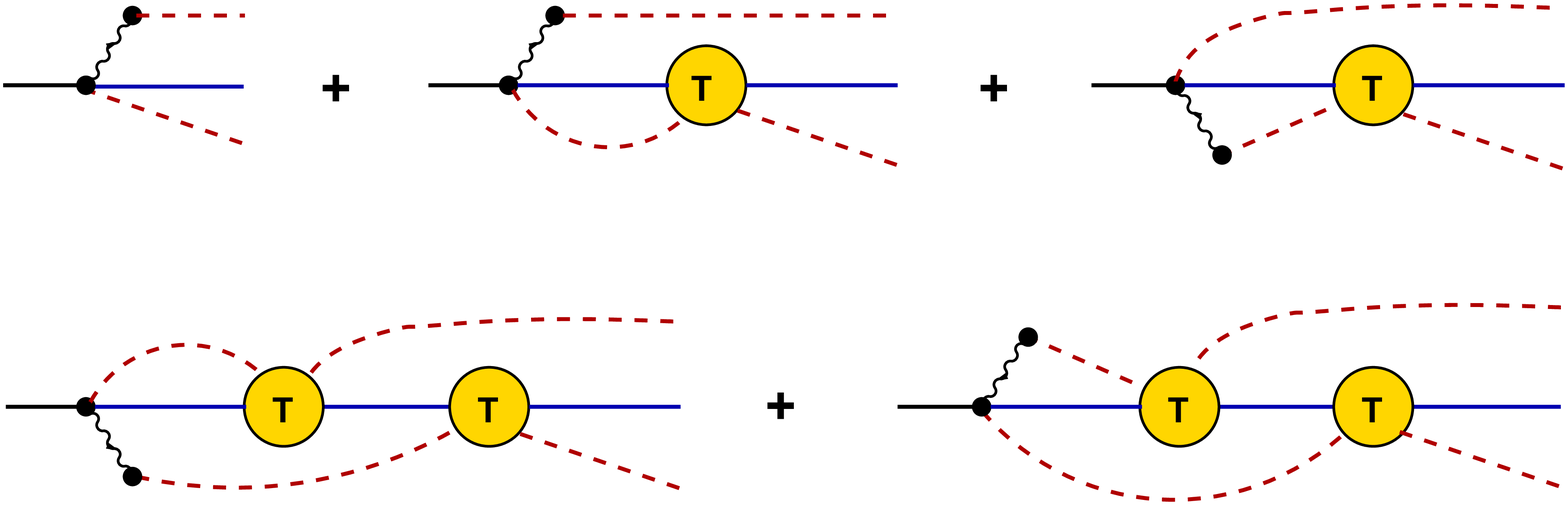}
\caption{Diagrams involving the weak vertex $W_a$;
the wavy line is a $W^+$, always plugged to a $\p^+$;
the $\p$ produced together with the $\Kb$ in the opposite
side can be either positive or neutral.}
\label{FPU-2}
\end{figure}

Processes involving the weak vertex $W_a$, defined in fig.\ref{FFSI-3},
are indicated in fig.\ref{FPU-2}.
The $W^+$ is shown explicitly just for the sake of clarifying the various topologies,
and is taken as a point in calculations.

 We start with the perturbative expansion given in eq.(\ref{pu.3}), 
 and we evaluate the one and two-loop terms as follows. 
 Using eqs.(\ref{deq6}) and (\ref{ea.8}), one gets
$\xi_1(m_{12}^2)=- W_a\,\Omega(m_{12}^2)$
and $\l_1(m_{12}^2)= - \, T(m_{12}^2)\, \Omega(m_{12}^2)$.
As $\Omega$ is divergent, a subtraction is needed and we assume
 the subtraction constant to be the same as in the $K\p$ 
 scattering amplitude. Using (\ref{ea.12}), one has
\bea
\l_1(m_{12}^2) &\!=\!& - \, T(m_{12}^2)\; [ C + \Ob(m_{12}^2)] \nn\\
&=& - 1 + \l'_1(m_{12}^2) \;,
\\[2mm]
\l'_1(m_{12}^2) &\!=\!& 1/D(m_{12}^2) \;.
\label{pu.4}
\eea
Our assumption  allows further simplifications in the
treatment of the rescattering series, but it is still a freedom
within our framework that we will not explore further, in order to
minimize the number of free parameters in our first investigation of
three-body rescattering effects.

The two-loop contribution comes from the recursive 
formula (\ref{deq6i}) and eq. (\ref{deq6ii}), and  reads
\bea
\l_2(m_{12}^2)
&\! =\!&  i\, T(m_{12}^2)\,
\int \! \frac{d^4 k}{(2\p)^4} \;
\frac{1}{[(p_{12} \sm k)^2 - M_\p^2 \sp i\,\e]\; }
\nn\\
&& \times \, \frac{1}{[k^2 - M_K^2 \sp i\,\e]}
\;T[(p_3 \sp k)^2] \;\Omega[(p_3\sp k)^2] \nn \\
\label{pu.5}
\eea
and, again, the divergent function $\Omega$ shows up.
Subtracting, we have
\bea
\l_2(m_{12}^2) &\!= \!& i\,T(m_{12}^2)\,
\int \! \frac{d^4 k}{(2\p)^4} \;
\frac{1}{[(p_{12} \sm k)^2 - M_\p^2 \sp i\,\e]\;  }
\nn\\
&& \times \,\frac{\;T[(p_3 \sp k)^2]}{
[k^2 - M_K^2 \sp i\,\e]} \;\lc C + \Ob[(p_3 \sp k)^2]\rc  .
\label{pu.6}
\eea
Using eq.(\ref{ea.12}) and subtracting once more, we get
\bea
\l_2(m_{12}^2) &\!= \!& \;T(m_{12}^2)\, [C+ \Ob(m_{12}^2)]
+ \l'_2(m_{12}^2) \nn\\
&=& -\l_1(m_{12}^2) + \l'_2(m_{12}^2) \;,
\\[4mm]
\l'_2(m_{12}^2) &\!= \!&
- \, i \, T(m_{12}^2) \int \! \frac{d^4 k}{(2\p)^4} \;
\frac{1}{[(p_{12} \sm k)^2 - M_\p^2 \sp i\,\e]}\;  \nn \\
&& \times \,\frac{1}{[k^2 - M_K^2 \sp i\,\e]}\;
\frac{1}{[D(p_3 \sp k)^2 ]}\;.
\label{pu.7}
\eea

Repeating this procedure for higher order loop contributions,
we find the structure
\beq
\l_{N}(m_{12}^2) = - \l_{N-1}(m_{12}^2) + \l'_{N}(m_{12}^2) \;,
\label{pu.8}
\eeq
which can be checked explicitly for  $\l_3$ and $\l_4\,$, using
the expressions shown in appendix \ref{perturbative contributions}.
Denoting by $S$ the sum of terms within the square bracket in
eq.(\ref{pu.3}), we have
\bea
S &=& \sum_{N=1}^\infty \lp \frac{2}{3}\rp^{N-1} \l_N \nn \\
&=& - \;\frac{2}{3} \,S -1
+ \sum_{N=1}^\infty \lp\frac{2}{3}\rp^{N-1}  \l'_N \, .
\label{pu.9}
\eea
This allows eq.(\ref{pu.3}) to expressed as
\bea
A_a(m_{12}^2,m_{23}^2) &\!=\!& \sqrt{\frac{2}{3}}\; W_a
\lb \l'_1(m_{12}^2) + \frac{2}{3} \; \l'_2(m_{12}^2)
 \right. \nn \\ &+& \left.
 \lp \frac{2}{3} \rp^2 \l'_3(m_{12}^2)
+ \cdots \rb + (1 \lrar 3).
\label{pu.10}
\eea
This result is interesting and indicates the importance
of treating both the $K\p$ amplitude and the FSIs in a single
coherent dynamical framework.
The diagrams of fig.\ref{FPU-2} contribute to the first two terms
of this series.

The integral (\ref{pu.7}) 
can be performed by a number of different techniques,
provided one recalls that $1/D(s)$ contains two poles,
associated with the $\k$ and the $K^*(1430)$.
One possibility would be to perform a Cauchy integration over $k_0$,
supplemented by a numerical three-dimensional integration.
An alternative, adopted here, is to rely on usual Feynman techniques.
With this purpose in mind, we employ the
expression for the $K\p$ amplitude written in terms of its poles,
given in eq.(\ref{ea.17}) and, using (\ref{pu.1}), obtain
\bea
\l'_2(m_{12}^2) &\!= \!&  T(m_{12}^2)
\lc - \b \;[C \sp \Ob(m_{12}^2)] \right.\nn\\ 
&+& \left.  \g_\k \P_{\p K \th_\k}/(16 \p^2)
+ \g_1  \P_{\p K \th_1}/(16 \p^2) \rc.
\label{pu.11}
\eea
The problem then reduces to the evaluation of triangle integrals.

\subsection{ Contributions proportional to $W_b$}

\begin{figure}[h]
\includegraphics[width=0.8\columnwidth,angle=0]{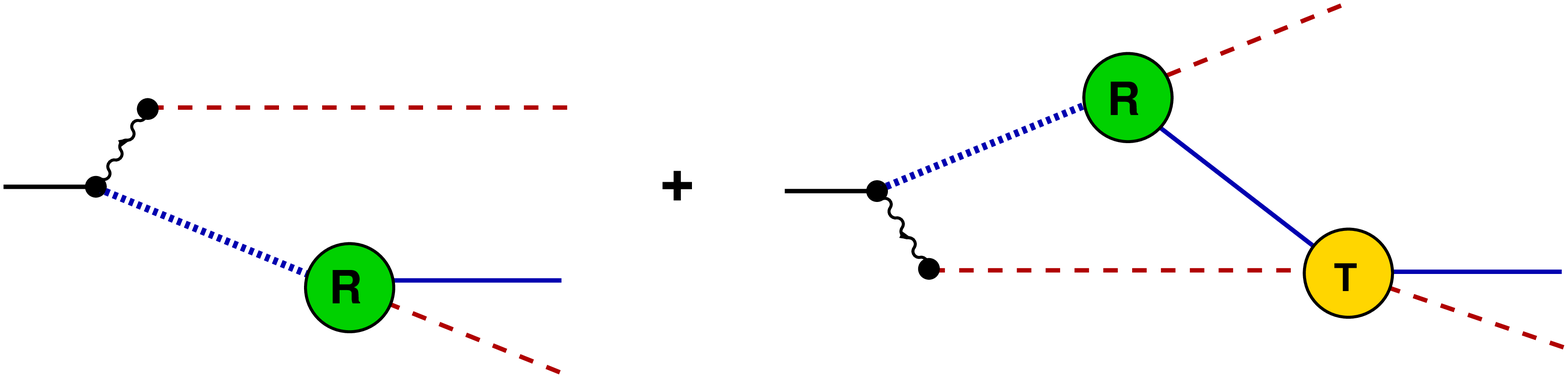}
\\
\vspace{10mm}
\includegraphics[width=0.9\columnwidth,angle=0]{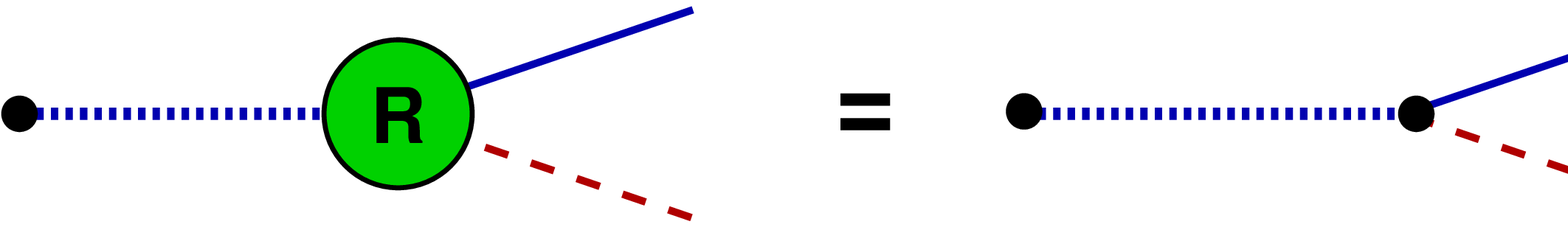}
\caption{Diagrams involving the weak vertex $W_b$;
the wavy line is a $W^+$, always plugged to a $\p^+$
and the dotted line is a scalar resonance,
which has a width given by the substructure $R$
described at the bottom line.}
\label{FPU-3}
\end{figure}

We follow here the treatment of the production amplitude for
a scalar resonance given in ref.\cite{Diogo}.
The basic idea is that processes, shown at the bottom line of
fig.\ref{FPU-3}, must always be considered together,
since the tree contribution in isolation contains a pole in the physical region.
The corresponding amplitude $R^i$, involving a resonance with mass
$m_i$, reads
\bea
R^i(m_{12}^2) &=& -i\; \frac{1}{F^2}
\lb c_d^i \,m_{12}^2 - (c_d^i \sm c_m^i)\,(M_\p^2 \sm M_K^2) \rb \,  \nn \\
&& \times \, \frac{1}{m_{12}^2 - m_i^2} \lb 1 - T(m_{12}^2)\,\Omega(m_{12}^2) \rb \;.
\label{pu.12}
\eea
Regularizing the function $\Omega$ and using eq.(\ref{ea.12})
one finds
\bea
R^i(m_{12}^2) &=& -i\;\frac{1}{F^2}
\lb c_d^i \,m_{12}^2 - (c_d^i \sm c_m^i)\,(M_\p^2 \sm M_K^2) \rb
\,  \nn \\
&&\times \,\frac{1}{[m_{12}^2 - m_i^2]\;D(m_{12}^2)} \;.
\label{pu.13}
\eea
This function is finite because, by construction, $D(m_i^2)=0$.

The evaluation of the contributions to $A_b$ is straightforward
and yields
\bea
A_b &\!=\!& a_{b1} + a_{b2} + \cdots \, , \\[2mm]
a_{b1}(m_{12}^2) &\!=\!& i \,W_b\; R^1(m_{12}^2) \;,
\label{pu.14}\\[2mm]
a_{b2}(m_{12}^2) &\!=\!& W_b\; \frac{2}{3}\; T(m_{12}^2) \,  \int \frac{d^4 k}{(2\p)^4} \;
\frac{R^1(p_3 \sp k)}{[(p_{12} \sm k)^2 - M_\p^2 \sp i\,\e]}\;  \nn\\
& \times &\,\frac{1}{[k^2 - M_K^2 \sp i\,\e]}\;, 
\label{pu.15}
\eea
where the recursive formula (\ref{deq6i}) is used starting  with $\xi_{1}(m_{12}^2)\equiv a_{b1}(m_{12}^2)$.
The function $a_{b2}$ can be reduced to a sum of triangle integrals
and is evaluated using eq.(\ref{pu.1}).

\subsection{ Contributions proportional to $W_c$}

\begin{figure}[h]
\includegraphics[width=1\columnwidth,angle=0]{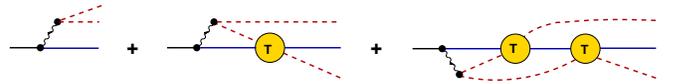}
\caption{Diagrams involving $W_c$; one of the pions
in the weak vertex is neutral.}
\label{FPU-4}
\end{figure}

Processes given  in fig.\ref{FPU-4} give rise to eq.(\ref{pu-c.3}), which contains the same

 series as in eq.(\ref{pu.3}).
Using result (\ref{pu.9}) one finds
\bea
A_c(m_{12}^2,m_{23}^2) &\!=\!& -\, {\frac{\sqrt{2}}{5}}\; W_c
\lb -1 + \l'_1(m_{12}^2) + \frac{2}{3} \; \l'_2(m_{12}^2)
\right.
\nn\\[2mm]
&\!+\!& \left.  \lp \frac{2}{3} \rp^2 \l'_3(m_{12}^2) \cdots \rb + (1 \lrar 3)\;.
\label{pu.17}
\eea
%

\section{ Results, conclusions and summary}

In this work we studied the role of final state  interactions in
$D^+ \rar K^- \p^+ \p^+$, treating two- and three-body interactions
in a consistent way.
Our underlying $K\p$  amplitude is derived from chiral perturbation
theory, supplemented by unitarization and tuned to elastic LASS data\cite{LASS}.
Two poles, associated with the $\k$ and the $K^*(1430)$, were then
 determined and employed in a representation of the amplitude.
A number of simplifying assumptions were  made in the $K\p$ amplitude,
for the sake of minimizing technical problems.
Among them, we mention the absence of both isospin $3/2$ and $P$ waves,
as well as couplings to vector mesons and to inelastic channels.
The means for correcting these shortcomings are available in the
literature and will be considered in future extensions of this work.
 Our treatment of three-body unitarity departs from a Faddeev-like
integral equation, which is subsequently expanded perturbatively.
Terms in the corresponding series contain a recursive component,
which allows a re-summation of the whole series,  when the divergence is subtracted using the
same criteria as in the $K\pi$ scattering amplitude.

 Model independent analyses\cite{brian,FOCUS,cleoc}
of the $s$-wave $K^-\p^+$ channel in the decay 
$D^+ \rar K^- \p^+ \p^+$ are rather welcome in theoretical studies,
because they are expressed in terms of amplitudes which are linear
in the Fermi constant.
This means that it is meaningful to study independently
the strong evolution of each dynamical contribution to the primary 
weak vertex.
In this work we have considered just three simple topologies of 
the color-allowed type to the primary vertex, which 
yield classes of decay amplitudes denoted by
$A_a\,$, $A_b\,$, and $A_c\,$
eqs.(\ref{pu.10}, \ref{pu.14}$,$\ref{pu.15}, \ref{pu.17}).
The first two begin at tree level and give rise to quasi two-body
FSIs, as in ref.\cite{DiogoRafael}.
The amplitude $A_c$, on the other hand, arises only when proper
three-body FSIs are present.

\begin{figure}[h]
\includegraphics[width=0.8\columnwidth,angle=0]{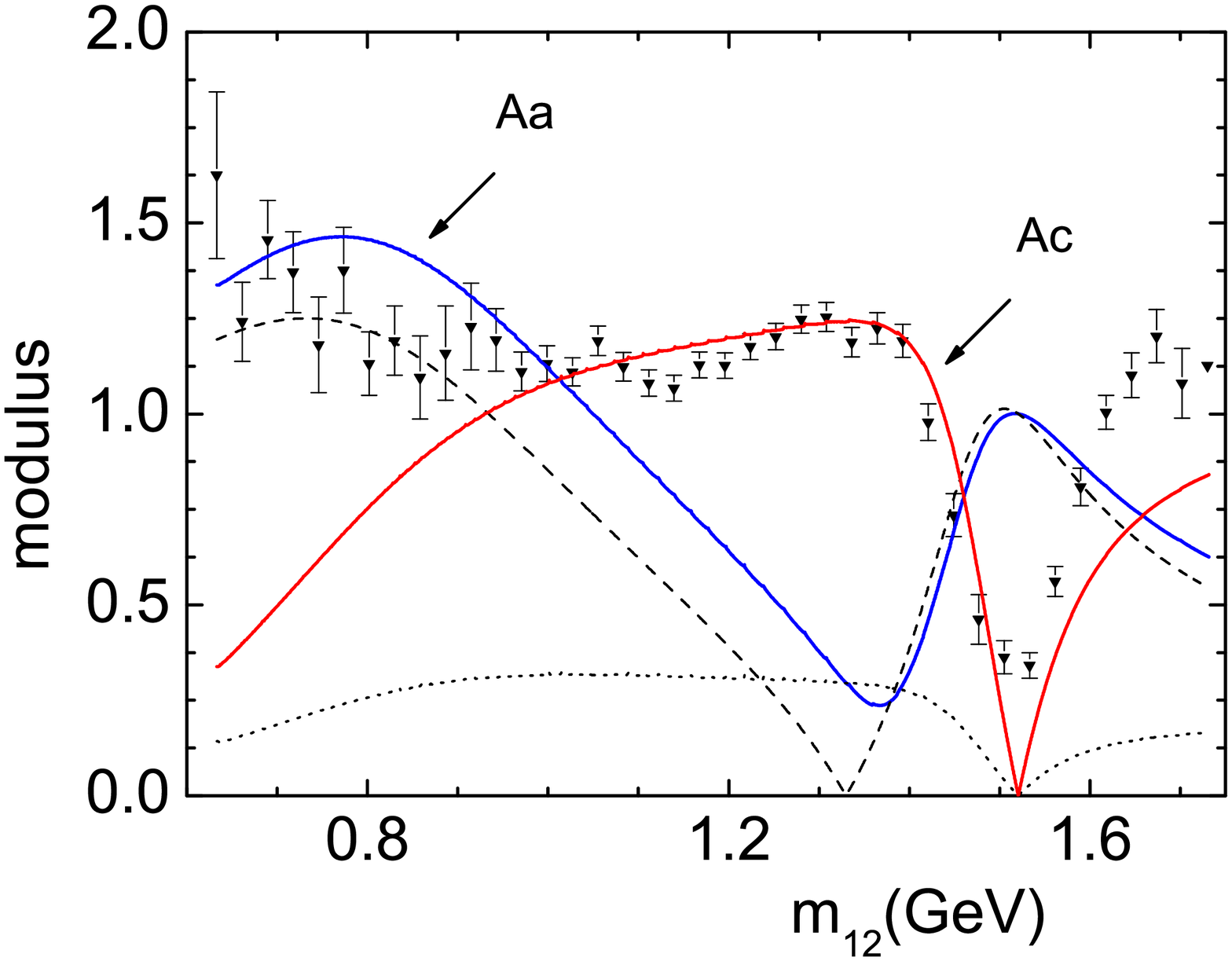}
\vspace*{-5mm}
\includegraphics[width=0.8\columnwidth,angle=0]{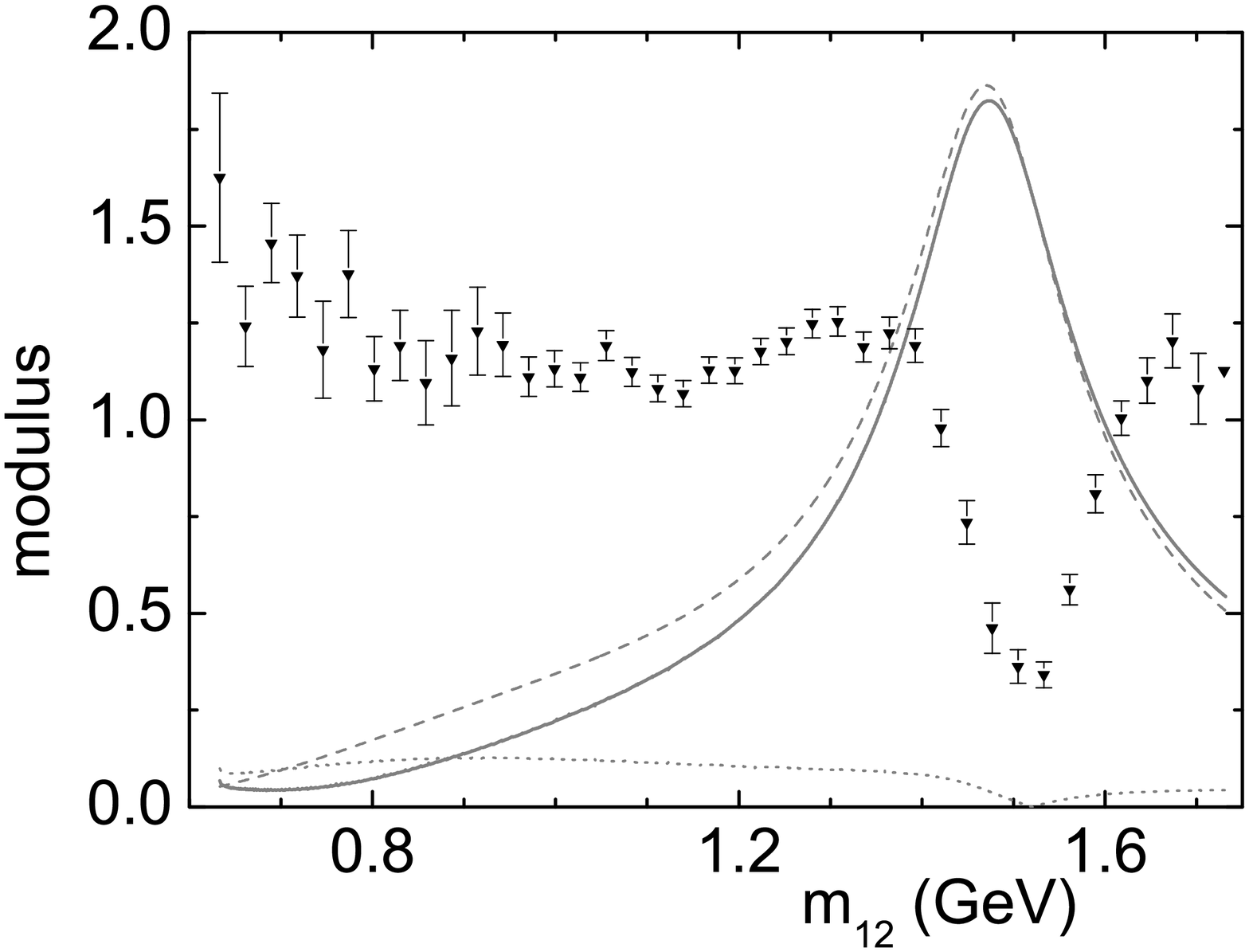}
\caption{up: behaviour of $|A_a|$ and $|A_c|$;
down: behaviour of $|A_b|$;
first and second order partial contributions are indicated by
dashed and dotted lines.}
\label{FPU-5}
\end{figure}

Predictions for their modulus are given in figs.\ref{FPU-5}.
It is important to note that the order $N$ of partial contributions
indicates the number of times the denominator $1/D$ of eq.(\ref{ea.12})
intervenes in a given function.
Results for $A_a$ and $A_c$, in particular,
are based on an infinite   re-summation of terms,
eq.(\ref{pu.9}), and hence $N$ is {\em not}
simply related with perturbative counting.
Understood in this sense, second order terms tend to be smaller that
leading ones.
However, in the case of $A_a$ and $A_c$, FSI corrections show up
clearly around the bare resonance mass,
where the function $1/D$ vanishes along the real axis.
In the case of $A_b$, one notices a cancellation
close to threshold.
As far as comparison with FOCUS data\cite{FOCUS} are concerned, we
note that $A_c$ has a  dip at the correct position,
whereas compatibility with the direct production of a resonance at the weak vertex is very difficult.

\begin{center}
\begin{figure}[h]
\includegraphics[width=0.95\columnwidth,angle=0]{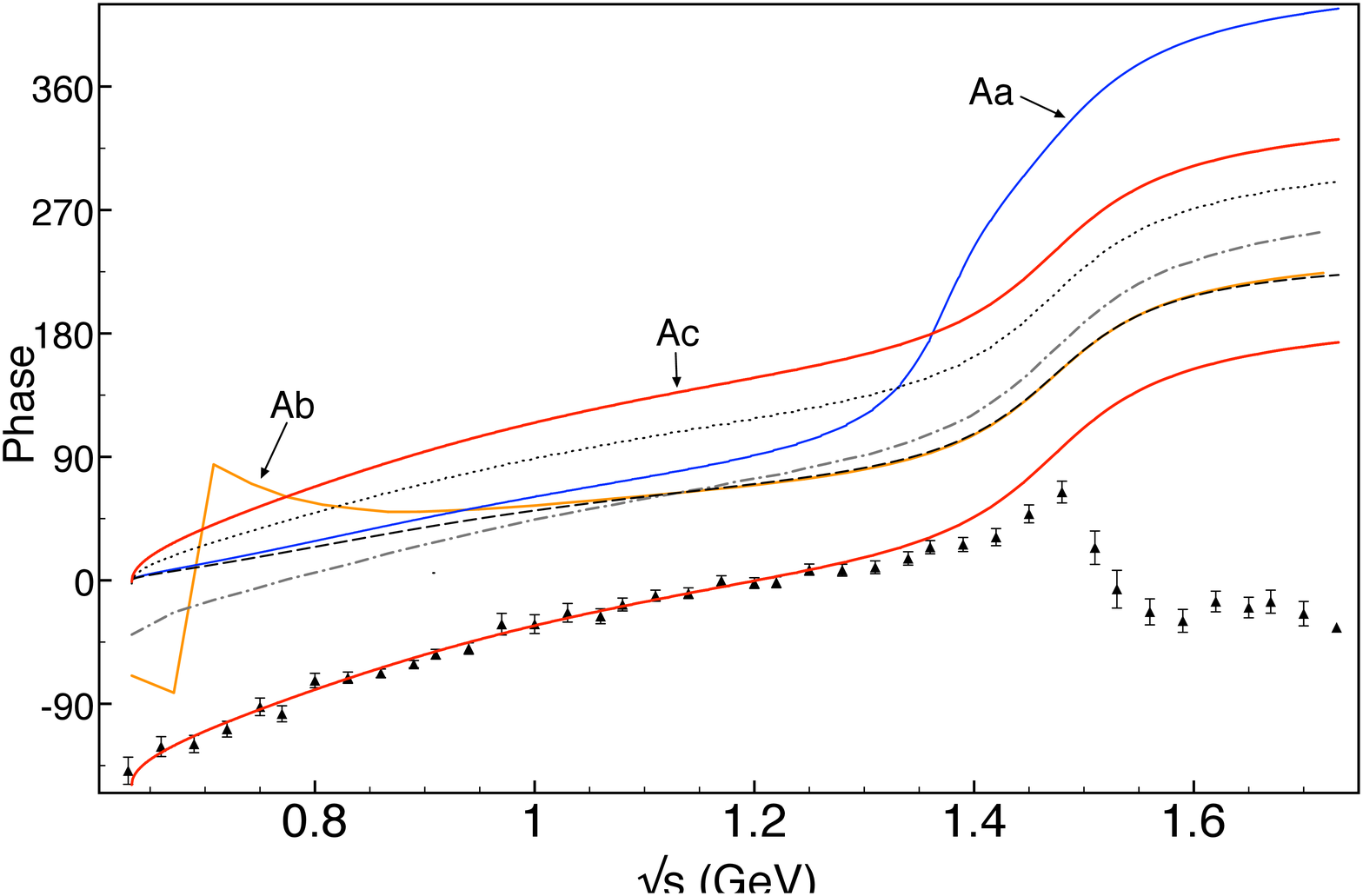}
\caption{Phases from $A_a\,$, $A_b\,$, and $A_c$, continuous line; partial contributions from $\l'_1$ in eqs.(\ref{pu.10}) and (\ref{pu.17}) and from $a_{b1}$ in eq.(\ref{pu.14}) coincide with the elastic  $K\pi$ phase, dashed lines; contributions from $\l'_2$ and $a_{b2}$ are given, respectively, by  dotted and dashed-dotted lines; the curve coinciding partially with FOCUS data\cite{FOCUS} (triangle) is $A_c$, shifted by $-148^0$.}
\label{FPU-6}
\end{figure}
\end{center}

Predictions for the phase are displayed in fig.\ref{FPU-6}.
Parcial contribution from $\l'_1$ in eq.(\ref{pu.10})
and eq.(\ref{pu.17}) and from $a_{b1}$ in eq.(\ref{pu.14}) fall exactly over the elastic $K\pi$ phase.  
The oscillation in $A_b$ at low-energies can be ascribed to the lack of precision in $a_{b2}$. 
We have checked that it is due to an incomplete cancellation between
a precise tree term and the less precise triangle contribution
in processes shown in fig.\ref{FPU-3}.
Finally, we see that a curve for $A_c$, shifted by $-148^0$,
describes well FOCUS data\cite{FOCUS} up to the region of the peak.

\begin{figure}[h]
\includegraphics[width=0.95\columnwidth,angle=0]{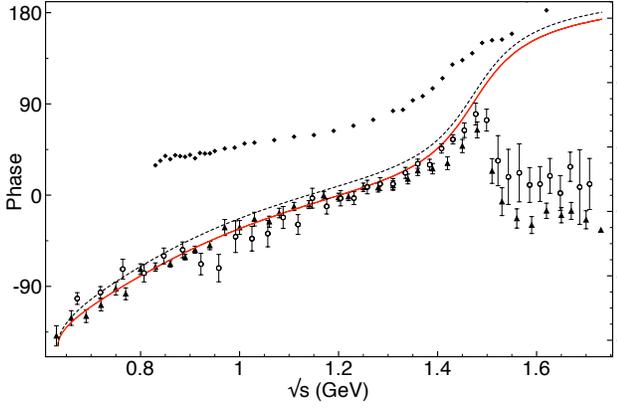}
\caption{Phases of the factors
$[-1 + \l'_1(m_{12}^2)]$ (dashed) and
$[-1 + \l'_1(m_{12}^2) + \frac{2}{3} \; \l'_2(m_{12}^2)]$ (continuous)
in eq.(\ref{pu.17}),
shifted by $-148^0$ compared with FOCUS \cite{FOCUS}(triangle) and E791\cite{brian}(circle) data, together with elastic $K\pi$ results from LASS\cite{LASS}(diamond).}
\label{FPU-7}
\end{figure}

As the study of both the modulus and the phase seems to favor the weak
vertex $W_c$ of fig.\ref{FFSI-3}, it is worth exploring its structure.
In fig.\ref{FPU-7}, we show the phases of the factors
$[-1 + \l'_1(m_{12}^2)]$ and
$[-1 + \l'_1(m_{12}^2) + \frac{2}{3} \; \l'_2(m_{12}^2)]$
in eq.(\ref{pu.17}).
The gap between them is about $10^0$ and the first term already
corresponds to a good approximation to low-energy FOCUS data.
However, as pointed out above, the factor $-1$ in these terms comes
from an infinite  re-summation and hence the smallness of this gap
{\em does not} indicate that it is enough to consider just
the first perturbative contribution.
Using eqs.(\ref{pu.17}), (\ref{pu.4}) and (\ref{ea.11}),
one has
 \bea
A_c(m_{12}^2,m_{23}^2) &\!\simeq \!& -\, {\frac{\sqrt{2}}{5}}\; W_c
\lb -1 + \frac{1}{1 + (C + \Ob) \,\cK }\rb \nn \\ && + (1 \lrar 3)\;,
\label{pu.18}
\eea
where $\cK$ is the $K\pi$ kernel of eq.(\ref{ea.4}).
Hence the dip in experimental results for the magnitude of $A$
indicates the energy at which $\cK=0$.
In our model, this corresponds to the point where both the real and
imaginary parts of the scattering amplitude cross the real axis
in fig.\ref{FFamplitude}.

In this framework, eq.(\ref{pu.18}), rewritten as
{\small \bea
A_c(m_{12}^2,m_{23}^2) &\!\simeq \!&  {\frac{\sqrt{2}}{5}}\; W_c
\lb \frac{(C + \Ob) \,\cK }{1 + (C + \Ob) \,\cK }\rb + (1 \lrar 3)\;,
\label{pu.19}
\eea}
could provide a much more convenient structure to be used in
experimental analyses, since $\Ob$ is a well known elementary function.

 As stressed before, this work concentrates at tracking 
leading effects, and a number of important issues were left untouched.
Among them, we mention the absence of $P$-waves, isospin $3/2$ 
terms and inelastic contributions to the basic $K\p$ amplitude,
as well as a more complete description of the primary weak vertex.
These matters will be dealt with in other papers.

In summary, our main conclusions read:
\\
{\em 1.} proper three-body effects are important;
\\
{\em 2.} corrections to the re-summed series are important at threshold
and fade away rapidly at higher energies;
\\
{\em 3.} a model based on a vector weak vertex can describe
qualitative features of the modulus from FOCUS and E791 data\cite{FOCUS, brian} for the decay 
amplitude
and agrees well with its phase in the elastic region.

\section*{Acknowledgments}
We thank D. R. Boito for important contributions to an earlier version of 
this paper.
This work was supported by Funda\c c\~ao de Amparo \`a Pesquisa do 
Estado de S\~ao Paulo (FAPESP) and Conselho Nacional de Desenvolvimento 
Cient\'\i fico e Tecnol\'ogico (CNPq) of Brazil.
\appendix

\section{two-meson propagator}
\label{bolha}

\begin{figure}[h]
\includegraphics[width=0.4\columnwidth,angle=0]{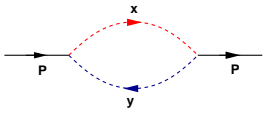}
\label{F11-a}
\caption{Bubble loop diagram.}
\end{figure}
%
The basic integral is
\bea
I_{\p K}(s) &\!=\!&  \int \frac{d^4\ell}{(2\p)^4} \;
\frac{1}{[(\ell \sp p/2)^2 - M_\p^2\,]\;[(\ell \sm p/2)^2 - M_K^2 \,]} \nn \\
&=& \frac{i}{(4\p)^2}\; \P_{xy}^{00}(s)\;,
\label{b.1}
\eea
since $P^2=s$.
In the Bethe-Salpeter equation, it is convenient to use
\beq
\Omega(s) \equiv i\, I_{\p K}(s)
\equiv
-  \lb L(s) \sp \Lambda_\infty \rb/16 \p^2\;,
\label{b.2}
\eeq
where $L(s)$ is a finite function and $\Lambda_\infty$ incorporates
the ultraviolet divergence.
The regular part of $\Omega$ is defined as
\bea
\Ob(s)= \Omega(s)-\Omega(0)
&=& - \lb L(s) \sm L(0) \rb/16 \p^2 \nn \\
&=& -  \Lb(s)/16 \p^2 \;,
\label{b.3}
\eea
where,
\bea
&\bullet & s \!<\! (M_\p\!-\!M_K)^2  
\nn \\
&&\Lb = \Lv
+ \frac{\sqrt{\l}}{s}\;
\ln \lb \frac{M_\p^2+M_K^2-s + \sqrt{\l}}
{2\,M_\p\,M_K}\rb \;,
\label{b.4}\\[3mm]
&\bullet &  (M_\p\!-\!M_K)^2 \!<\! s\! < \!(M_\p^2\!+\!M_K^2) 
\nn \\ 
&&\Lb= \Lv
- \frac{\sqrt{- \l}}{s}\;
\tan^{-1} \lb \frac{\sqrt{-\l}} {M_\p^2+M_K^2-s}\rb \;,
\label{b.5}\\[3mm]
&\bullet & (M_\p^2\!+\!M_K^2) \!<\! s \! < \! (M_\p\!+\!M_K)^2 
\nn \\
&&\Lb =  \Lv
- \frac{\sqrt{- \l}}{s}\;
\lc \tan^{-1} \lb \frac{\sqrt{-\l}}
{M_\p^2+M_K^2-s}\rb + \p \rc \;,
\label{b.6}\\[3mm]
&\bullet & s \! > \! (M_\p\!+\!M_K)^2 
\nn \\
&&\Lb =  \Lv
- \,\frac{\sqrt{\l}}{s}\;
\ln \lb \frac{s- M_\p^2- M_K^2 + \sqrt{\l}}
{2\,M_\p\,M_K}\rb \nn\\
&&\,\,\,\,\,\,\,+\,\, \,i\,\p\;\frac{\sqrt{\l}}{s} \;,
\label{b.7}
\eea
with
\bea
 \Lv &= & 1 + \frac{M_\p^2 + M_K^2}{M_\p^2 - M_K^2}\;\ln \lb\frac{M_\p}{M_K}\rb
-  
\frac{M_\p^2-M_K^2}{s}\;\ln \lb \frac{M_\p}{M_K} \rb \;, \nn \\
\label{b.8}\\[3mm]
&& \l = s^2 - 2\;s\;(M_\p^2+M_K^2) + (M_\p^2-M_K^2)^2 \;.
\label{b.9}
\eea

\section{Isospin in FSIs}
\label{isospin}

In order to derive the isospin weighting factors presented in the
integral equation (\ref{deq5i}), one has to realize that the adopted
model for the $K\pi$ transition matrix has a separable form, which
allows for the factorization of the decay amplitude as given by eq.
(\ref{deq2}). In addition, we consider only the dominant isospin 1/2
channel of the $K\pi$ system. The isospin 1/2 dependence in the
transition matrix is made explicit by writing it in the form:
\begin{equation} T_{1/2}\equiv \sum_{i_z}|I_{K\pi}=1/2, i_z\rangle\; T\;
\langle I_{K\pi}=1/2, i_z| \ , \label{i1}\end{equation} where the
matrix element of $T$ is given by eq. (\ref{ea.17}).

The $K\pi\pi$ scattering amplitude with $I_z=3/2$ is built only with
total isospin states $I_T=3/2$, because the $K\pi$ subsystem
interacts only in the isospin doublet channel. The states with
isospin $5/2$ do not contribute to the final state interaction in
our model. The above simplification implies that the spectator
function $\xi$, which also carries the bachelor pion momentum
distribution, is given by:
\bea |\xi_{3/2}(k_\pi)\rangle=|I_T, I_z,
I_\pi,I_{K\pi^\prime}\rangle \;
 \xi(k_\pi) \ , \label{i2}
\eea
where $I_T=3/2$, $I_z=3/2$, $I_{\pi}=I_{\pi^\prime}=1$ and
$I_{K\pi}=I_{K\pi^\prime}=1/2$.

Following the diagrammatic representation of the integral equation
(\ref{deq5i}) shown in fig. \ref{FFSI-4}, one realizes that the kaon
is exchanged between the two possible interacting $K\pi$ pairs.
Guided by the diagram, one verifies that the bachelor pion in the
intermediate state interacts with the kaon and forms a new pair.
This is a known property of the Faddeev decomposition, which equates
only different Faddeev components of the 3$\to$3 transition matrix.

The kernel of the integral equation exchanges the role of the
bachelor pions from the intermediate to the final state (see fig.
\ref{FFSI-4}). In addition, by considering the I=1/2 transition
amplitude (\ref{i1}) and the isospin structure of the spectator
function (\ref{i2}), one finds that the weighting isospin factor
multiplying the kernel of eq. (\ref{deq5i}) is given by the
re-coupling coefficient
\begin{equation} R_{3/2}\equiv \langle I_T, I_z, I_\pi,I_{K\pi^\prime} |
I_T, I_z, I_{\pi^\prime},I_{K\pi}\rangle \ , \label{recou2}
\end{equation}
which is written in terms of Wigner 6-j symbol, \bea
R_{3/2}=(-1)^S f(I_{K\pi})f(I_{K\pi^\prime})\left\{%
\begin{array}{lll}
I_\pi & I_K & I_{K\pi}\\
I_\pi & I_T & I_{K\pi^\prime} \\
\end{array}
\right\}, \label{6j} \eea with $S=2I_\pi+I_K+I_T$ and
$f(I)=\sqrt{2I+1}$. The result is the weight factor $R_{3/2}=2/3$ in
(\ref{deq5i}).

In the particular case of eq. (\ref{deq5i}), the driving term for
$I_T=3/2$  is derived from a weak vertex symmetric under the
exchange of the pions, such that:
\begin{equation} |W_{3/2}\rangle \!=\! W \left( |I_T, I_z,
I_\pi,I_{K\pi^\prime}\rangle \! + \! |I_T, I_z,
I_{\pi^\prime},I_{K\pi}\rangle\right) \ . \label{w32}\end{equation}
Both terms contribute to the driving term isospin weighting factor
after projection onto the appropriate isospin state, which
corresponds to the spectator function (\ref{i2}). One gets that
\bea &&\langle I_T, I_z, I_\pi,I_{K\pi^\prime} |W_{3/2}\rangle =
\nn \\
 &&= W\langle I_T, I_z, I_\pi,I_{K\pi^\prime}|I_T, I_z,
I_\pi,I_{K\pi^\prime}\rangle +
\nonumber \\
 &&+ W \langle I_T, I_z, I_\pi,I_{K\pi^\prime} |
 I_T, I_z, I_{\pi^\prime},I_{K\pi}\rangle
=\frac53 W \ , \label{recou1} \eea
which was computed by using the isospin re-coupling coefficient
(\ref{6j}).


\section{ Perturbative contributions}
\label{perturbative contributions}

Perturbative contributions to the  FSI amplitude
$a(m_{12}^2)=\sum_{N=1}^\infty a_{N}(m_{12}^2)\;$, eq.(\ref{tbu.2}), have the structure
\beq
a_N(m_{12}^2) = - a_{N-1}(m_{12}^2) + a_1(m_{12}^2) \;\l_N(m_{12}^2)
\label{pc.1}
\eeq
and, in the sequence, we list the $\l'_{N}(m_{12}^2)$ for $N=3, 4$.
\bea
\l'_3(m_{12}^2) &\!=\!& -T(m_{12}^2)
\int \frac{d^4 k}{(4\p)^4}\;
\frac{1}{(p_{12} \sm k)^2 \sm M_\p^2\sp i\,\e} \; 
\nn\\[2mm]
&\! \times \!&
\ \frac{1}{k^2 \sm M_K^2 \sp i\,\e}\;T(k \sp p_3)
\nn\\[2mm]
&\! \times \!& \int \frac{d^4 k'}{(4\p)^4}\;
\frac{1}{[(k \sp p_3)\sm k'\,]^2 \sm M_\p^2 \sp i\,\e} \;  \nn\\ 
&\! \times \!& \frac{1}{k^{'2} \sm M_K^2 \sp i\,\e}\; \frac{1}{D[(k' \sp p_{12} \sm k)^2]} \;,
\label{pc.4}\\[4mm]
\l'_4(m_{12}^2) &\!=\!& i\,T(m_{12}^2)
\int \frac{d^4 k}{(4\p)^4}\;
\frac{1}{(p_{12} \sm k)^2 \sm M_\p^2 \sp i\,\e} \;\nn\\ 
&\! \times \!&  \frac{1}{k^2 \sm M_K^2 \sp i\,\e}\;T(k \sp p_3)
\nn\\[2mm]
&\! \times \!& \int \frac{d^4 k'}{(4\p)^4}\;
\frac{1}{[(k \sp p_3)\sm k'\,]^2 \sm M_\p^2 \sp i\,\e}  \;
\nn\\ 
&\! \times \!& 
\frac{1}{k^{'2} \sm M_K^2 \sp i\,\e}\;T[(p_{12} \sm k) \sp k'\,]
\nn\\[2mm]
&\! \times \!& \int \frac{d^4 k''}{(4\p)^4}\;
\frac{1}{\{[k' \sp (p_{12} \sm k)] \sm k'' \,]\}^2 \sm M_\p^2 \sp i\,\e} \;
\nn\\ 
&\! \times \!& 
\frac{1}{k^{''2} \sm M_K^2 \sp i\,\e}\; \frac{1}{D[(p_3 \sp k \sm k' \sp k'')^ 2]}\;. \nn \\
\label{pc.5}
\eea

\section{Triangle integral}
\label{TrI}

The triangle integral defined by eq.(\ref{pu.1}) and represented
in fig.\ref{FPU-1} is written as
\bea
I_{\p K \th} &\!=\!& \frac{i}{(4\p)^2}\,\P_{\p K \th} \,,\nn\\ 
&\! = \!& 
 - \frac{i}{(4\p)^2}\,
\int_0^1 da\; a \, \int_0^1 db\; \frac{1}{D_{\p K \th}}\;,
\label{tri.1}\\[4mm]
D_{\p K \th} &\!=\!& (1 \sm a)\,M_\p^2 + a\,(1 \sm b)\,M_K^2 + a\,b\; \th_R
\nn\\[2mm]
&\! - \!& i\,[\,\e \sp a\,b\,(\th_I-\e)\,]
- a\,(1 \sm a)(1 \sm b)\,m_{12}^2
\nn\\[2mm]
&\! - \!& a\,(1 \sm a)\,b\,M_D^2 - a^2 b\,(1 \sm b)\,M_\p^2 \;.
\label{tri.2}
\eea
The double integral (\ref{tri.1}) can be evaluated numerically but,
in the case $\th_I = \e$, problems of accuracy may arise.
In order to understand the structure of its complex part,
it is desirable to pursue the analytic path as long as possible.
We therefore resort to the $SU(2)$ chiral limit and neglect $M_\p^2$,
eliminating terms quadratic in $b$ in $D_{SK\p}$ and
simplifying the algebra.
Integration in $b$ yields
\bea
J &\! = \!& \int_0^1 db\; \frac{a}{D_{\p K \th}}
\nn\\[4mm]
&\! = \!& \frac{G \sp i\,(\th_I \sm \e)}{G^2 + (\th_I \sm \e)^2}
\lc \frac{1}{2}\, \ln
\frac{[F \sp G]^2 \sp [(\th_I \sm \e)+ \e/a]^2}
{F^2 \sp (\e/a)^2} \right.
\nn\\[4mm]
&\! - \!& \left.
i\,\frac{F(\th_I \sm \e) - G\,(\e/a)}
{|F(\th_I \sm \e) - G\,(\e/a)|}  \;
\lb \tan^{-1} \frac{x \sm y}{1 \sp x\,y} + \s \; \p \;\rb\rc \;,
\nn\\[4mm]
F &\! = \!& M_K^2 - (1 \sm a)\,m_{12}^2 \;,
\nn\\[2mm]
G &\! = \!& (\th_R \sm M_K^2) - (1 \sm a) \, (M_D^2 \sm m_{12}^2)  \;,
\nn\\[4mm]
x &\!=\!& \frac{(F \sp G)\,G \sp (\th_I \sm \e)\,(\th_I \sm \e \sp \e/a)}
{|F(\th_I \sm \e) - G\,(\e/a)|} \;,
\nn\\[4mm]
y &\!=\!& \frac{F \, G \sp (\th_I \sm \e) \,(\e/a)}
{|F(\th_I \sm \e) - G\,(\e/a)|}\;,
\label{tri.3}
\eea
with $[xy>-1] \rar \s=0\;$, $[xy<-1$ and $x>0] \rar \s=+1\;$,
$[xy<-1 $ and $ x<0] \rar \s=-1\;$.
This result was used to tune numerical calculations and one
found that convergence for small values of $\e$  requires 
a large number of points in a Gaussian integration.


\end{document}